\documentclass[aps,pra,twocolumn,superscriptaddress,longbibliography,nofootinbib]{revtex4-2}

\usepackage{graphicx}
\usepackage{dcolumn}
\usepackage{braket}
\usepackage{bm}
\usepackage{booktabs}
\usepackage{graphicx}
\usepackage{enumitem}
\usepackage{amsmath}
\usepackage{amssymb}
\usepackage{hyperref}
\usepackage[utf8]{inputenc}
\usepackage{mathtools}
\usepackage[english]{babel}
\usepackage{bbm}
\hypersetup{colorlinks=true, linkcolor=blue, citecolor=blue, urlcolor=blue}
\usepackage{xcolor}
\usepackage{braket}
\usepackage[normalem]{ulem}
\DeclareMathAlphabet{\mathbbold}{U}{bbold}{m}{n}
\usepackage{bbm}
\usepackage{chngcntr}
\usepackage{apptools}
\usepackage{float}

\AtAppendix{%
  \counterwithin{figure}{section}
  \counterwithin{table}{section}

}
\usepackage{url}

\begin{document}

\title{Light-induced Self-Organization in Cooperative Free Space Atomic Arrays}

\author{Sara Molló-Guri}
\affiliation{Department of Physics, Harvard University, Cambridge, Massachusetts 02138, USA}

\author{Oriol Rubies-Bigorda}
\affiliation{Department of Physics, Harvard University, Cambridge, Massachusetts 02138, USA}
\affiliation{Physics Department, Massachusetts Institute of Technology, Cambridge, Massachusetts 02139, USA}

\author{Raphael Holzinger}
\affiliation{Department of Physics, Harvard University, Cambridge, Massachusetts 02138, USA}

\author{Jonah S. Peter}
\affiliation{Department of Physics, Harvard University, Cambridge, Massachusetts 02138, USA}
\affiliation{Department of Chemistry and Chemical Biology, Harvard University, Cambridge, Massachusetts 02138, USA}

\author{Susanne F. Yelin}
\affiliation{Department of Physics, Harvard University, Cambridge, Massachusetts 02138, USA}

\begin{abstract}

We investigate how laser-driven, cooperative dipole–dipole interactions in weakly trapped atomic arrays give rise to self-organized configurations.
Starting from an analytically tractable two-emitter system, we identify the possible steady-state spatial arrangements accessible to the atoms. We then extend this analysis to larger ensembles in both linear and ring geometries. In linear chains, we demonstrate the emergence of topologically nontrivial dimerized configurations across a range of initial interatomic spacings. In ring geometries, we find that the system undergoes self-organized contraction and expansion, enabling access to length scales below those set by the trapping lattice. 
Our results demonstrate that collective light–matter interactions in free space can spontaneously generate modified ordered geometries, even when the emitters are initially separated by distances larger than their transition wavelength.
\end{abstract}

\maketitle

\setlength{\parindent}{10pt}
\setlength{\parskip}{4pt}
\section{\label{sec:level1}Introduction}

Atomic self-organization refers to the reordering of atoms by light-mediated forces, with the ensemble evolving towards a configuration of minimal energy. This behavior emerges from the coupling between the internal degrees of freedom of the atoms and their motional dynamics~\cite{WALLIS1995203,Dalibard_1985,Chu1985, Dalibard1989}, and can give rise to a wide variety of complex phenomena. In cavity quantum electrodynamics (QED), for example, collective self-organization facilitates cavity-mediated cooling, enables long-range interatomic interactions, and can induce non-equilibrium phase transitions\cite{Domokos2002, Black2003, Baumann2010,Domokos2003,asboth2005self,black2003observation,schutz2015thermodynamics}). Similar physics has been explored in waveguide QED~\cite{chang2013selforganization,eldredge2016self}, where atoms can spontaneously form spatial patterns with an emergent lattice constant determined by light-matter interactions. 

In recent years, there has been a renewed interest in studying collective light-matter interactions in free space. In closely spaced atomic ensembles, collective dipole-dipole interactions give rise to cooperative phenomena such as superradiance and subradiance~\cite{dicke_coherence_1954, bonifacio_quantum_1971, bonifacio_cooperative_1975, harochesuperradiance, AsenjoGarcia2017}, resulting in pronounced modifications to the system's optical response. 
In ordered atomic arrays, precise control over the lattice geometry enables the emergence of nontrivial topological states, which are further enriched by cooperative decay processes~\cite{Perczel_topology,peter_chirality_2024, peter_chirality-induced_2024, ring_raphael}. The collective optical responses of atomic arrays can also be harnessed to control atomic motion~\cite{SHAHMOON20191,shahmoon2020quantum, BuckleyBonanno2024}, leading to applications like collectively-enhanced laser cooling~\cite{maximo2018optical,oriolraphaelcooling} and self-organization in free space~\cite{ho2025optomechanical,maximo2018optical}. Moreover, recent advances in experimental techniques—particularly optical lattices~\cite{Rui2020,Greiner_superradiance} and tweezer arrays~\cite{HUANG2023100470,Barredo2016,Endres2016}—provide a powerful platform for studying these collective effects.
These systems enable atom-by-atom assembly of defect-free arrays with tunable geometries and nearest-neighbor spacings on the order of, or even below, the atomic transition wavelength. Combined with single-atom resolution and precise control over the internal quantum states, this level of programmability offers an ideal setting for exploring self-organization driven by light-mediated forces in previously inaccessible regimes.

\begin{figure}[t] 
    \centering
    \includegraphics[width=\linewidth]{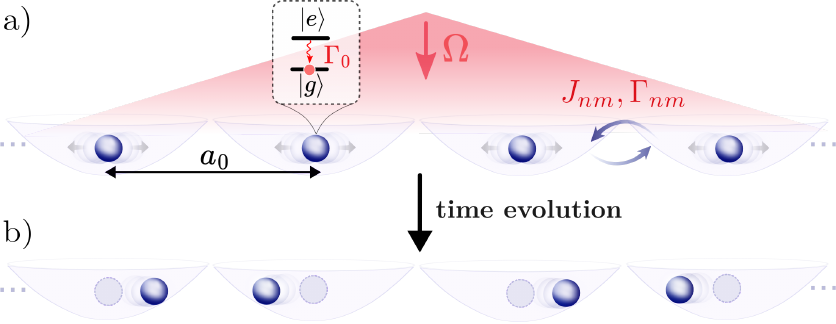} %
    \caption{(a) Schematic of an ordered array of $N$ two-level quantum emitters with an initial nearest-neighbor separation $a_0$ set by the trap center, which is of the order of the transition wavelength $\lambda_0= 2\pi c/\omega_0$.  
    The atoms are confined in shallow harmonic traps with frequency $\omega\ll\Gamma_0$, where $\Gamma_0$ denotes their spontaneous decay rate. 
    Dipole interactions are induced by an external laser drive (Rabi frequency $\Omega$) that acts uniformly on all atoms. Collective resonances emerge from the long-range coherent (dissipative) dipole-dipole interactions $J_{nm}$ ($\Gamma_{nm}$). (b) The interplay of weak harmonic confinement, continuous laser driving, and strong dipole-dipole interactions leads to the self-organization of atoms into stationary spatial configurations. As we show in this work, dimerized atomic chains constitute one of the possible self-organized configurations.}
    \label{fig:schematics}
\end{figure}

Here, we show that light-induced dipole–dipole interactions in free space atomic arrays give rise to emergent order that can be tuned by varying the system's initial parameters. 
We develop a theoretical framework for 
atomic emitters arranged in one-dimensional chains and ring geometries, confined by weak harmonic traps and coherently driven by laser light [Fig.~\ref{fig:schematics}]. In linear chains, we find that light-induced dimerization emerges across a broad range of lattice spacings, resulting in the spontaneous formation of nontrivial topological states assembled through collective dipole-dipole interactions. In ring geometries, we uncover a mechanism for self-organized radial contraction and expansion, enabling access to length scales below those set by the trapping potential. These findings advance the understanding of light-induced collective motion in free space and open new avenues for engineering dynamical states of matter through cooperative dipole-dipole forces.

\section{\label{sec:model}Model}
We consider $N$ two-level atoms, each having electronic energy states $\ket{g}$ and $\ket{e}$ with resonance frequency $\omega_0$ and spontaneous decay rate $\Gamma_0$. We assume the atoms are tightly confined in the \textit{z} direction, but are individually trapped within weak harmonic potentials of frequency $\omega$ in the \textit{x-y} plane, thus allowing for in-plane lateral motion [Fig.~\ref{fig:schematics}a)]. We treat the atom-laser interaction in the semiclassical regime, where the atoms are coherently driven by a plane wave laser of frequency $\omega_L = ck_L$, detuning $\delta=\omega_L-\omega_0$ and Rabi frequency $\Omega= \mathbf{d}^*\cdot \mathbf{E}_0/{\hbar}$, where ${\bf d}$ is the dipole matrix element of the atomic transition (assumed to be identical for each atom and independent of the trapping transition). The positive frequency component of the classical electric field of the laser at position $\mathbf{r}$ is expressed as $\mathbf{E}^+(\mathbf{r})=\mathbf{E}_0 \mathrm{exp}[{i( \mathbf{k}_L\cdot \mathbf{r}-\omega_Lt)}]$. The atoms interact via vacuum-mediated dipole-dipole interactions, which in turn induce mechanical motion on the atoms themselves. After tracing out the vacuum degrees of freedom in the Born-Markov limit~\cite{lehmberg} and applying the rotating wave approximation to the laser drive, the Hamiltonian of the atomic system reads $\hat{H}=\hat{H}_{\text{laser}}+\hat{H}_{\text{dipole}}+\hat{H}_{\text{motion}}$, with
\begin{align}
\hat{H}_{\text{laser}} &= -\hbar  \Big[  \sum_{n=1}^N \Omega\hat{\sigma}^{\dagger}_n e^{i(\mathbf{k}_L \cdot \hat{\mathbf{r}}_n-\omega_L t)}+ \text{h.c.} \Big]\nonumber \\
\hat{H}_{\text{dipole}} &=\hbar\omega_0\sum_{n=1}^N\hat{\sigma}_n^{\dagger}\hat{\sigma}_n+\hbar\sum_{n,m}^N C_{nm}\hat{\sigma}_n^\dagger\hat{\sigma}_m  \\
\hat{H}_{\text{motion}} &= \sum_{n=1}^N\sum_{i=x,y} \bigg[ \frac{(\hat{p}_{ni})^2}{2m}+\frac{m\omega^2 {\hat{r}}_{ni}^2}{2} \bigg]. \nonumber
\label{eq:hams}
\end{align}

Here, $\hat{\sigma}_n=\ket{g_n}\bra{e_n}$ is the lowering operator for atom \textit{n}, $\hat{r}_{ni}$ and $\hat{p}_{ni}$ are the $i$th component of the atomic position and linear momentum, respectively, and h.c. stands for Hermitian conjugate. The quantity $C_{nm}$ describes the dipole-dipole interaction between atoms \textit{n} and \textit{m}, and is defined as
\begin{align}
C_{nm}=J_{nm}-i\frac{\Gamma_{nm}}{2}=-\frac{3\pi\Gamma_0}{\omega_0} \mathbf{d}^\dagger\mathbf{G}(\mathbf{r}_{nm},\omega_0)\mathbf{d},
\end{align}
where $\Gamma_0$ is the single-atom spontaneous emission rate and the quantities $J_{nm}$ and $\Gamma_{nm}$ correspond to the coherent and dissipative coupling rates of the dipole-dipole interaction, respectively. These rates are determined by the Green's tensor $\mathbf{G}(\mathbf{r}_{nm}, \omega_0)$ for electric dipoles in free space, where $\mathbf{r}_{nm} = \mathbf{r}_{n} - \mathbf{r}_{m}$ is the vector pointing from atom \textit{m} to \textit{n} (see Appendix~\ref{app:greentensor}).

In the frame rotating at the laser frequency, the equations of motion for the atomic coherences, defined as $\hat{\tilde{\sigma}}_n=\hat{\sigma}_n e^{i\omega_Lt}$, are governed by the Heisenberg–Langevin equations. In the weak-driving regime, $\Omega \ll \Gamma_0$, the system remains in the low-excitation limit, such that $\langle \hat{\sigma}^{\dagger}_n\hat{\sigma}_n\rangle-\langle\hat{\sigma}_n\hat{\sigma}^{\dagger}_n\rangle  \approx -1$ at all times (see Appendix~\ref{app:refdim}). In this regime, the spin degrees of freedom are fully characterized by the mean-field coherences $\tilde{\sigma}_n \equiv \langle\hat{\tilde{\sigma}}_n\rangle$, which evolve according to
\begin{equation}
\label{eq:sigmadot}
\frac{d}{dt}\tilde{\sigma}_n = \left(i\delta - \frac{\Gamma_0}{2}\right)\tilde{\sigma}_n  -i 
   \bigg( \sum_{m \ne n} C_{nm}\tilde{\sigma}_m - \Omega \bigg).
\end{equation}
Here, we have assumed a planar atomic array driven by a normally incident laser [Fig.~\ref{fig:schematics}(a)], such that $\mathbf{k}_L\cdot\mathbf{r}=0$, and therefore all atoms experience the same phase imparted by the driving laser.

Similarly, the mean-field dynamical equations for the atomic positions, $r_{ni} \equiv \langle \hat{r}_{ni}\rangle$, and momenta, $p_{ni} \equiv \langle \hat{p}_{ni}\rangle$, are given by
\begin{align}
\label{eq:rdot}
\dot{r}_{ni}&=\frac{ p_{ni}}{m}
\end{align}
\begin{align}
\label{eq:pdot}
\dot{p}_{ni}&=-\hbar \sum_{m \ne n} \bigg[ \frac{d C_{nm}}{d r_{ni}} \tilde{\sigma}_n^*  \tilde{\sigma}_m + \text{h.c.} \bigg] - m\omega^2(r_{ni} - r_{t_n}^i),
\end{align}
where $r_{t_n}^i$ denotes the center of the $n$-th harmonic trap (\textit{i.e.}, the equilibrium position in the absence of inter-atomic interactions). 

In the regime of weak coherent driving and slow atomic motion ($\omega_r \ll \Gamma_0$), the internal optical degrees of freedom relax on a timescale of $\sim \Gamma_0^{-1}$, which is much faster than the mechanical dynamics. This separation of timescales allows us to treat the internal dynamics as adiabatically following the atomic positions.
In this limit, light-mediated interactions effectively reshape the potential energy landscape, resulting in local minima that determine the new equilibrium configurations. We obtain these new equilibrium positions by treating the motional degrees of freedom at the mean-field level. While a full quantum mechanical treatment would describe fluctuations around these positions, here we focus on locating the equilibrium configuration itself. 

In our numerical treatment, we solve the coupled system of Eqs.~(\ref{eq:sigmadot}-\ref{eq:pdot}), with each atom initialized at rest in its electronic ground state at the trapping position $\mathbf{r}_n = \mathbf{r}_{t_n}$. We define the system to be \emph{self-organized} when the atomic positions converge to a stationary configuration [Fig.~\ref{fig:schematics}b)], corresponding to a force-balanced fixed point of Eqs.~(\ref{eq:rdot})--(\ref{eq:pdot}), or equivalently, a local minimum of the adiabatic effective potential. To facilitate efficient convergence to a nearby local minimum, we include a small auxiliary damping term $-\gamma p_{ni}$ in Eq.~(\ref{eq:pdot}). This phenomenological friction coefficient accounts for additional dissipative processes~\cite{shahmoon2020quantum,chang2013selforganization}, which generally depend on the experimental implementation. In our simulations, we take $\gamma=0.005\Gamma_0$, which ensures convergence to a steady-state, and $\omega_r/\Gamma_0 = 10^{-3}$, where $\omega_r = \hbar k_0^2/m$ ($k_0 \! = \! \omega_0/c$) is the recoil frequency of each atom.

\section{\label{sec:two}Two-emitter case}

The physics of self-organization is most easily understood for the case of two atoms, where the equations of motion admit a simple analytical treatment. We consider two atoms moving along the \textit{x}-axis while driven by a laser propagating along \textit{z}.

\begin{figure}[t]
\includegraphics[width=\linewidth]{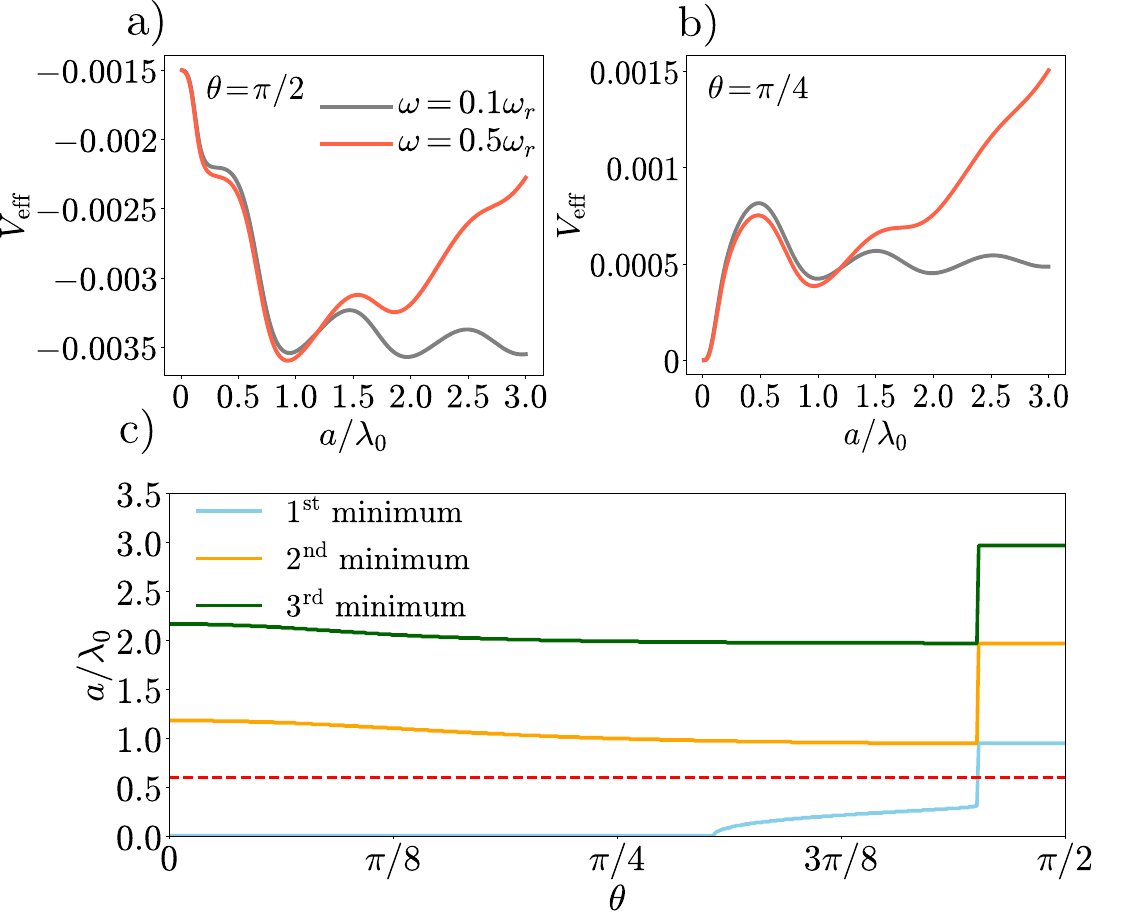}
\caption{Steady-state behavior for two atoms with identical dipole moments $\mathbf{d}=[\cos\theta,i \sin\theta,0]$ 
and electric field polarized along $\mathbf{d}$. Effective potential for (a) $\theta=\pi/2$ and (b) $\theta=\pi/4$ as a function of the relative position $a$ between the two atoms. (c) Local minima of the effective potential $V_\mathrm{eff}$ with $\omega=0.1\omega_r$ as a function of $\theta$, corresponding to atomic separations at which the atoms are stationary and self-organized.
The red dashed line denotes the initial separation $a_0$ of the atoms at their trap equilibrium positions. All panels were obtained for a coherent driving rate $\Omega=0.05\Gamma_0$, on resonance with the atomic frequency and with trap positions separated by $a_0 =0.6\lambda_0$.}
\label{fig:eff}
\end{figure}

In the limit $\omega_r \ll \Gamma_0 $, we can adiabatically eliminate the internal degrees of freedom to obtain an analytical solution (see Appendix~\ref{app:refdim}). By setting $\langle \dot{\tilde{\sigma}}_n \rangle =0$ in Eq.~(\ref{eq:sigmadot}) we obtain the quasi-static steady-state values of $\langle\tilde{\sigma}_n\rangle$. The amplitudes of the collective symmetric and anti-symmetric spin operators, $\langle \tilde{\sigma}_S \rangle = (\langle \tilde{\sigma}_1\rangle + \langle \tilde{\sigma}_2\rangle)/\sqrt{2}$ and $\langle \tilde{\sigma}_A\rangle = (\langle \tilde{\sigma}_1\rangle - \langle \tilde{\sigma}_2\rangle)/\sqrt{2}$, under the adiabatic elimination and in the weak excitation limit are $\langle\tilde{\sigma}_S\rangle = \sqrt{2}\Omega/(C_{12}-\delta-i\frac{\Gamma_0}{2})$ and $\langle\tilde{\sigma}_A\rangle = 0$.

Due to the symmetry of the two-atom system, the center-of-mass and relative motional degrees of freedom decouple. The center-of-mass coordinate $x_{\mathrm{CM}} = (x_1 + x_2)/2$ and momentum $p_{\mathrm{CM}} = (p_1 + p_2)/2$ evolve as a simple harmonic oscillator about the trap midpoint $\bar{x}_{\mathrm{trap}} = (x_{t_1} + x_{t_2})/2$, with equations of motion $\dot{p}_{\mathrm{CM}} = -m\omega^2 (x_{\mathrm{CM}} - \bar{x}_{\mathrm{trap}})$ and $\dot{x}_{\mathrm{CM}} = p_{\mathrm{CM}}/m$. For atoms initially at rest at their trap centers, such that $x_{\mathrm{CM}}(0) = \bar{x}_{\mathrm{trap}}$ and $p_{\mathrm{CM}}(0) = 0$, the center-of-mass remains stationary, and the dynamics are completely described by the relative degrees of freedom: the interatomic separation $a = x_2 - x_1$ and the relative momentum $p_{\mathrm{R}} = (p_2 - p_1)/2$. Their equations of motion are obtained from Eq.~(\ref{eq:pdot}) as 

\begin{align}
  \dot{a}=\frac{p_\text{R}}{m},
  \label{adot}
\end{align}
\begin{align}
\dot{p}_\text{R} = \frac{-2\hbar \left|\Omega\right|^2}
{(C_{12}-\delta)^2 + \Gamma_0^2/4} 
\frac{dJ_{12}}{da}
- \frac{1}{2} m \omega^2 (a - x_{t_2}),
\label{pdotr}
\end{align}
where we have set $x_{t_1}=0$ without loss of generality. 
The steady-state solutions of Eqs.~(\ref{adot}) and (\ref{pdotr}) correspond to configurations in which the atoms self-organize at the minima of the effective potential $V_\mathrm{eff}=-\int \dot{p}_R \ da$, yielding a modified interatomic separation $a$.

Fig.~\ref{fig:eff}a) and b) show the computed potentials as a function of relative atomic separation $a$ for two dipole moment orientations $\mathbf{d}=[\cos\theta,i \sin\theta,0]$, with $\theta=\pi/2$ and $\theta=\pi/4$, respectively. The electric field is polarized along the same direction as $\mathbf{d}$, such that it maximizes the laser-atom interaction. Each local minimum corresponds to a stable steady-state separation at which the atoms can self-organize. 
The first three local minima are shown in Fig.~\ref{fig:eff}c) as a function of the dipole orientation angle $\theta$.
The red dashed line indicates the initial atomic separation, corresponding to both atoms initially positioned at the centers of their respective traps (here chosen as $a_0 \equiv x_{t_2} - x_{t_1} = 0.6\lambda_0$).
Depending on the value of $\theta$, the atoms can self-organize at separations smaller (first minimum in the range $\theta \in [0.3,0.45\pi]$) or larger than the initial distance. For dipole angles $\theta \in [0,0.3\pi]$, the effective potential exhibits a minimum at $a=0$, indicating that the interaction becomes attractive at short distances, leading to collisions between the atoms and unstable behavior.

To validate the analytical treatment based on adiabatic elimination described above, we numerically integrate the full equations of motion and confirm that the atoms relax into one of the minima shown in Fig.~\ref{fig:eff}. The specific minimum reached depends on the initial conditions (i.e.\  the initial position and momentum of the atoms) and on the friction coefficient, as reduced friction allows the atoms to retain enough kinetic energy to overcome local barriers and bypass the nearest minimum.

\section{\label{sec:chain}Linear chain}

In this section, we show how the self-organization of a linear atomic chain can lead to the formation of topological states via light-mediated dimerization of the constituent atoms. We consider chains arranged along the $x$-axis, with atoms moving along $x$ and dipole moments oriented along $z$. The laser propagates along the \textit{y}-axis and its electric field is polarized along \textit{z}. The emergence of dimerization for this setup can be intuitively understood by extending the two-atom behavior to larger ensembles. For the parameters of Fig.~\ref{fig:eff}a), the effective potential is negative, implying that interactions between any two atoms are repulsive. In a three-atom chain, the central atom experiences equal repulsion from its two neighbors and therefore remains stationary, while the outer atoms move outward. For a four-atom chain, the outer atoms repel their inner neighbors, pushing the two middle atoms toward each other and thereby forming a dimer. This mechanism naturally results in the formation of a quasi-periodic dimerized chain for larger system sizes. However, because the interatomic interactions strongly depend on the initial lattice spacing and on the detuning of the driving laser, complete dimerization does not occur in all regimes, as shown below.


In order to quantify the extent to which a self organized chain becomes dimerized, we define the \textit{dimer strength} $D_s$ as
\begin{eqnarray}
    D_s=\frac{1}{N-2}\sum_{n=1}^{N-2}(-1)^n\frac{|J_{n+1,n+2}|-|J_{n,n+1}|}{|J_{n+1,n+2}|+|J_{n,n+1}|}.
    \label{eq:ds}
\end{eqnarray}

Each term in the sum compares consecutive interatomic couplings through a normalized difference of the nearest-neighbor dipole-dipole interaction $J$, while the prefactor $(-1)^n$ ensures constructive contributions for chains with alternating atomic spacings. In other words, dimerized chains yield a finite $D_s$, whereas a uniform chain with a single lattice constant gives $D_s = 0$. The magnitude $|D_s|$ quantifies the degree of dimerization, with larger values corresponding to a larger difference between the two alternating spacings. Additionally, the sign of $D_s$ distinguishes between two configurations: $D_s>0$ corresponds to a fully dimerized chain (where all atoms are paired), and $D_s<0$ to a dimerized chain with unpaired atoms at the edges. As an illustrative example, for alternating couplings $J_s$ (strong, intra-dimer) and $J_w$ (weak, inter-dimer), Eq.~(\ref{eq:ds}) gives $D_s = (J_s - J_w)/(J_s + J_w) > 0$ for a fully dimerized chain, and $D_s = (J_w - J_s)/(J_s + J_w) < 0$ when the edge atoms are unpaired.

\begin{figure}[hb]
\includegraphics[width=0.4\textwidth]{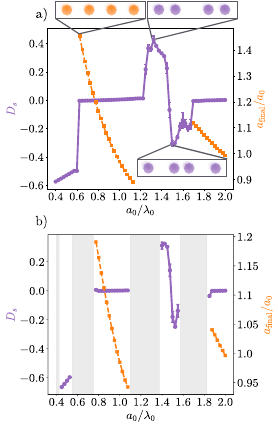}
\caption{\label{fig:dim} Dimer strength $D_s$ (purple circles) for a chain of (a) $N=4$ atoms and (b) $N=10$ atoms, with dipole and laser polarizations orientated along $z$. $D_s=0$ indicates a uniformly ordered chain, $D_s>0$ corresponds to a fully dimerized chain, and $D_s<0$ to a dimerized chain with two unpaired atoms at the edges. The insets illustrate schematic representations of the self-organized configurations in each case. Larger values of $\vert D_s \vert$ indicate a stronger contrast between the two alternating interatomic distances.
For $D_s\approx 0$, orange squares show the final interatomic spacing of the chain normalized to its initial value ($a_\mathrm{final}/a_0$). The shaded regions correspond to parameter regimes in which the final configurations are neither dimerized nor uniformly ordered chains. In both panels, we consider $\omega=1.0\omega_r$, $\Omega=0.05\Gamma_0$ and $\delta=0$. Symbols correspond to average results over 30 realizations with uniformly distributed random positional disorder in the initial trap positions within the interval $[-0.01\,a_0,\; 0.01\,a_0]$. Error bars denote standard deviation.}
\end{figure}

Fig.~\ref{fig:dim}a) shows the dimer strength for a four-atom chain as a function of the initial lattice spacing $a_0$ and with the driving laser tuned on resonance with the atomic transition (see Appendix~\ref{app:others} for results with non-zero detunings). Dimerized configurations emerge over a broad range of initial spacings. However, for specific values of $a_0$, the dimer strength approaches zero, indicating that the array forms an ordered chain with a single modified lattice spacing (orange curves in Fig.~\ref{fig:dim}).

Increasing the number of atoms reveals similar behavior as for $N=4$, although more complex phenomena can also emerge due to the growing complexity of all-to-all interactions. For $N=10$ [Fig.~\ref{fig:dim}b)], we again observe both the formation of dimerized chains and chains with modified lattice spacings depending on the initial configuration. These phases are interspersed by configurations that cannot be readily classified as either a dimerized or an equidistant chain [shaded regions in Fig.~\ref{fig:dim}b], as they lack two well-defined consecutive, alternating interatomic distances (see Appendix~\ref{app:others}).


\subsection*{Topological edge states}
The simplest model that exhibits topological edge states is the Su-Schrieffer-Heeger model (SSH), which describes a fully Hermitian dimerized chain with nearest-neighbor coupling~\cite{su1979solitons}. For ultracold atoms coupled to free space modes, the system is inherently more complex, involving long-range dipole-dipole interactions and an effective non-Hermiticity. Moreover, the atomic positions in the self-organized configurations described above deviate slightly from those of an idealized periodic dimerized chain. However, despite these deviations, topologically protected edge modes can persist~\cite{nonhermtop}, and we observe these modes in our self organized chains.

  \begin{figure}[hbt]
\includegraphics[width=\linewidth]{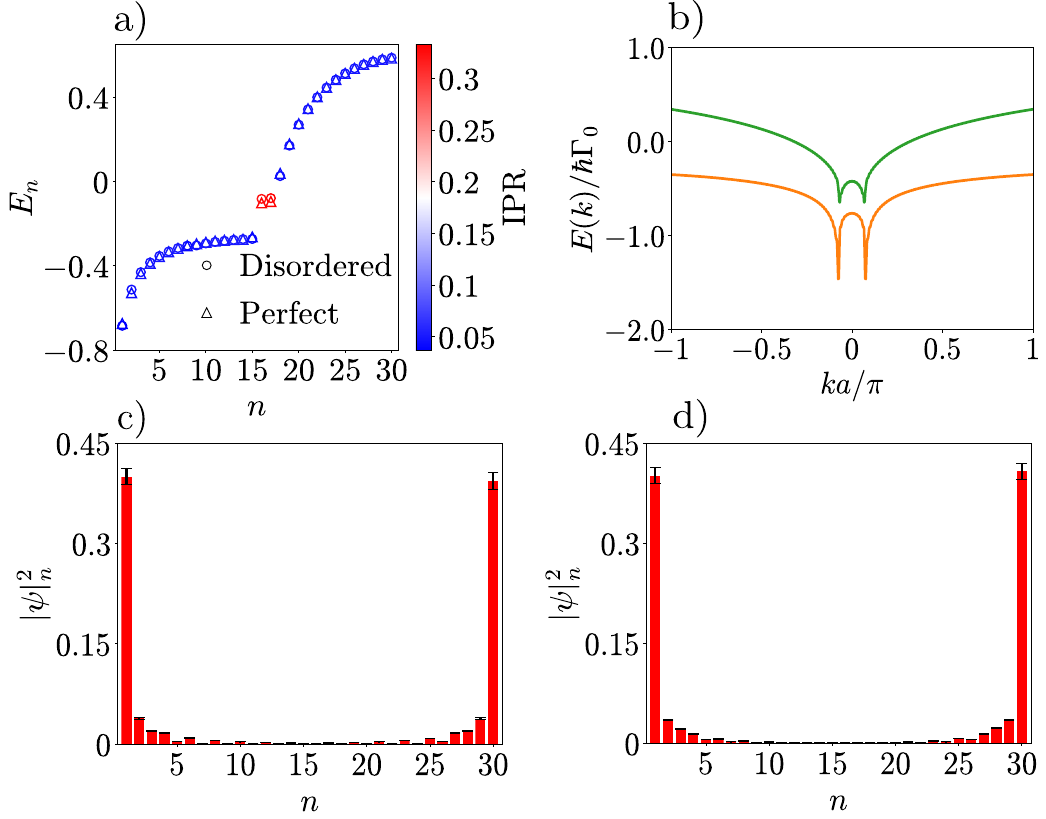}
\caption{\label{fig:edgestates} a) Real part of the eigenvalue spectrum of the non-Hermitian Hamiltonians $H_\mathrm{eff}$ for self-organized (circles) and perfectly dimerized (triangles) chains. Circles correspond to a single realization of a chain of $N=30$ atoms with initial trap spacing $a_0=0.5\lambda$. Initial disorder in the trap positions is uniformly distributed in $[-0.01\,a_0,\; 0.01\,a_0]$, resulting in a self-organized dimerized chain with slight variations in the alternating separations. The spectrum of the most similar perfectly dimerized configuration (see main text) largely overlaps with that of the self-organized chain. The inverse participation ratio [IPR, Eq.~(\ref{eq:ipr})] is maximal for two nearly-degenerate eigenstates, indicating localization at the edges of the chain. b) Band structure for a perfectly dimerized chain, with the two lattice constants given by the mean separations of the self-organized chain. c,d) Spatial probability distributions of the quasi-degenerate states, highlighting their localization at the system edges. The error bars show the variance over 30 realizations of the initial trap position fluctuations. We consider $\omega=1.0\omega_r$, $\Omega=0.05\Gamma_0$ and $\delta=0$.  }
\end{figure}

Here, we investigate chains of $N=30$ atoms with an initial trap spacing $a_0 = 0.5\lambda_0$, and with additional uniformly random positional disorder. For each realization, we solve the equations of motion to obtain a self-organized dimerized configuration that exhibits small deviations from a perfectly periodic dimerized chain. These fluctuations are characterized by a standard deviation of the alternating separations of approximately $s \sim 10^{-2}/\lambda_0$, which remains much smaller than the atomic separations themselves.
Fig.~\ref{fig:edgestates}a) shows the eigenstates of the atomic system for a representative realization, obtained by diagonalizing the effective non-Hermitian Hamiltonian  $H_\mathrm{eff}=\sum_{n,m}^N C_{nm}\tilde{\sigma}^{\dagger}_n\tilde{\sigma}_m$ for the self-organized chain. Its eigenvalue spectrum closely matches that of the perfectly dimerized chain whose two alternating bond lengths are taken as the mean distances extracted from the self-organized configuration. These similarities suggest that the disorder in the final positions does not significantly alter the system’s properties. Notably, the two middle states are approximately degenerate and their probability distribution [Fig.~\ref{fig:edgestates}c),d)] is localized at both ends of the chain. We quantify the degree of localization in these edge states via the inverse participation ratio (IPR)~\cite{ipr}:
\begin{equation}
    \label{eq:ipr}IPR=\frac{\sum_n^N|\psi_n|^4}{[\sum_n^N|\psi_n|^2]^2},
\end{equation}
where $\psi_n$ denotes the amplitude of eigenstate $\psi$ on site $n$. Localized edge states exhibit IPR values much larger than those of bulk states. This contrast is clearly visible in the color map of Fig.~\ref{fig:edgestates}a), where bulk states appear in blue (small IPR) and edge states in red (large IPR).

To verify that these edge states are topologically nontrivial, we compute the Zak phase, defined as $\varphi = i \int_{\mathrm{BZ}} dk\, \braket{\psi_k | \partial_k \psi_k}$,
for the perfectly periodic dimerized chain described above. This restores translational invariance and serves as a figure of merit for the disordered chain, which itself does not possess exact periodicity. The resulting band structure indeed exhibits a non-zero Zak phase, confirming the presence of nontrivial topology. The small disorder in the self-organized chain leads to a maximum absolute difference between the corresponding matrix elements of the self-organized and periodic Hamiltonians of $\sim 0.03\Gamma_0$. This discrepancy is significantly smaller than the minimum band gap of the perfectly periodic chain [$\sim 0.49\Gamma_0$, Fig.~\ref{fig:edgestates}b)], suggesting that the disorder present in the self-organized chain is unlikely to close the gap, and therefore the edge states remain topologically protected.

\section{\label{sec:ring}Ring Geometry}

\begin{figure}[ht]
\includegraphics[width=\linewidth]{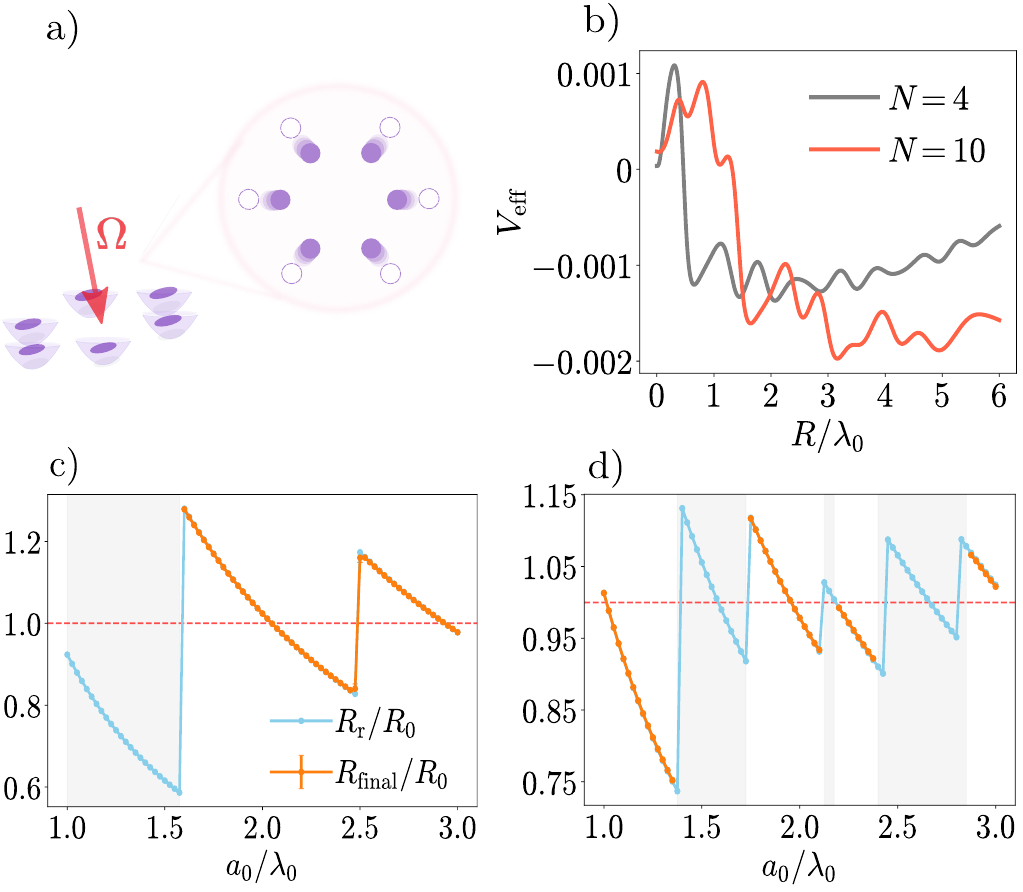}
\caption{\label{fig:ring}
a) Schematic of the trapped ring configuration showing gradual contraction of the ring radius over time. The dipoles are circularly polarized within the plane of the ring and the atoms are driven by a laser propagating perpendicular to that plane. b) Effective potential in the radial direction of the ring for different atom numbers, with initial interatomic distance $a_0=1.5\lambda_0$. c-d) Radii of the self-organized ring geometries for (c) $N=4$ and (d) $N=10$ as a function of initial interatomic distance. The blue curves ($R_{\mathrm{r}}/R_0$) correspond to radial simulations without initial position disorder, while the orange curves ($R_{\mathrm{final}}/R_0$) show the results of full simulations with $1\%$ positional disorder in the initial $x$ and $y$ positions, averaged over 30 runs. Error bars are smaller than the symbol size. The gray regions denote initial configurations that lose their ring structure during the relaxation towards a stationary configuration. For $a_0 <\lambda_0$, the motional dynamics can become unstable. In all panels, we consider $\omega=0.1\omega_r$, $\Omega=0.05\Gamma_0$ and $\delta=0$. Note that the trap frequency is lower than that used for the linear chains to allow more pronounced radial variations while avoiding dimer collisions. } 
\end{figure}

In this section, we study self-ordering in arrays of $N$ emitters uniformly placed on a ring of radius $R$, such that the nearest-neighbor spacing is given by $a=2R\sin(\pi/N)$. Ring geometries are not only realizable with current experiments~\cite{HUANG2023100470}, but appear abundantly in natural photosynthetic complexes~\cite{Koepke1996,Ritsch_subradiance_9fold}. When the dipole moments are circularly polarized in the plane of the ring and the atomic system is driven transversely, as illustrated in Fig.~\ref{fig:ring}a), the permutational symmetry of the ring geometry is preserved. Under these conditions, all atoms are equivalent ($\tilde{\sigma}_n \equiv\tilde{\sigma}$ $\forall n$) and their motion is restricted to the radial direction [Fig.~\ref{fig:ring}a)], effectively reducing the problem to a single particle moving along a single direction. Applying the adiabatic elimination introduced in previous sections, we obtain the equations of motion   
\begin{align} \label{eq:ring}
\tilde{\sigma} &=\frac{-i\Omega}{i\delta-\Gamma_0/2-i\sum_{m = 2}^N C_{1m}}, \\ 
    \dot{p} &=-2\hbar|\tilde{\sigma}|^2\sum_{m =2}^N \frac{dJ_{1m}}{dR}-m\omega^2(R-R_t),  
\end{align}
and $\dot{R}=p/m$. Here, $R$ denotes the radius of the ring (that is, the radial distance of each atom from the center) and $p$ is the associated radial momentum (identical for all atoms). In addition, $R_t$ denotes the radius corresponding to the center of each atomic trap.
Similar to the two-atom case, we construct the effective potential as $V_{eff}= -\int \dot{p} \ dR$, shown in Fig.~\ref{fig:ring}b) for a representative configuration. The resulting landscape features multiple local minima, each corresponding to a stable steady-state radius. Crucially, when the initial radius is smaller than that of the first equilibrium configuration, the optical force pushes the atoms inward, causing the ring to contract until it is halted by the repulsive forces of the effective potential.

Simulations of the radial motion without initial positional disorder yield the self-organized radii $R_{\mathrm{r}}/R_0$, shown as blue curves in Fig.~\ref{fig:ring}c),d), which exhibit both expansion and contraction of the ring by up to $25\%$ of the initial radius. To assess the stability of these configurations, we next introduced in-plane positional disorder in the initial atomic positions. This disorder breaks the permutation symmetry, and therefore the motion is no longer constrained to be purely radial. The resulting steady-state radii, $R_{\mathrm{final}}/R_0$, obtained by solving the full equations of motion are shown in orange.
While many configurations coincide with the disorder-free case, the shaded regions indicates parameter regimes in which the ring structure becomes unstable, either due to random distortions or the formation of atomic pairs. This behavior highlights that certain initial interatomic spacings lead to particular instabilities that disrupt the ring geometry.

\section{Discussion and Conclusions} \label{sec:conclusions} 
We have investigated the collective motion of atoms in free space arising from a combination of laser-induced dipole-dipole interactions and weak harmonic confinement. Our results indicate that atomic self-organization is realizable under experimentally-relevant conditions and that that these effects persist even when the interatomic spacing exceeds the resonant wavelength. The minimal two-atom case already reveals the existence of multiple stable spatial arrangements and could be implemented with modern optical platforms. Extending to larger systems, linear chains of emitters exhibit the emergence of dimerized patterns for certain initial spacings, while for others the lattice undergoes an overall expansion or contraction. 
In the subwavelength regime, these dimerized configurations can host topologically protected edge states, opening a route toward dynamical control of topological phases via external laser parameters. Our results also lay the groundwork for future studies investigating the interplay between cooperative light emission and self-organization. Both subwavelength atomic chains and ring geometries support highly localized superradiant and subradiant states~\cite{ring_raphael}. As such, our results may provide a mechanism to tailor collective emission properties for applications in quantum light generation~\cite{St-Jean2017LasingTopologicalEdgeStates}.

A central assumption of the adiabatic treatment used in this work is the separation of timescales between fast internal (optical) and slow external (motional) degrees of freedom. For many dipole-allowed optical transitions, $\Gamma_0/2\pi$ lies in the MHz range, whereas trap frequencies in optical tweezer arrays or optical lattices are typically in the kHz--100\,kHz range. These transitions ensure that the condition $\omega/\Gamma_0 \ll 1$ is satisfied. The complementary requirement that the zero-point motion remains small compared to the lattice spacing is discussed in Appendix~\ref{app:zpm}.

An experimental realization of this setup could be achieved using programmable arrays of optical tweezers~\cite{Pause2024laser,holman2024trappingsingleatomsmetasurface} or optical lattices~\cite{Baier2016,Du2024_dipolar_bilayer,Greiner_superradiance}, which enable the deterministic arrangement of atoms into ring or one-dimensional chain geometries with high spatial precision~\cite{brow2017optical,glicenstein2020collective,HUANG2023100470}. Subwavelength optical lattices based on short-wavelength transitions offer a viable platform for realizing large atomic ensembles, particularly with alkaline-earth atoms~\cite{Du2024_dipolar_bilayer,Baier2016,Rui2020,Rubies_thesis_mit,Greiner_superradiance}. Alternatively, optical tweezer arrays employ positioning techniques capable of arranging tens to thousands of atoms with high accuracy and single-site reconfigurability~\cite{Seubert2025, Norcia2018,Pause2024laser}, making them especially well-suited for engineering the geometries considered in this work.

\vspace{1em}
\noindent \emph{Acknowledgments -} We are grateful to C. Rusconi for
insightful discussions. S.F.Y. acknowledges NSF via the CUA PFC PHY-2317134, and QuSeC-TAQS OMA-2326787 in addition to AFOSR FA9550-24-1-0311. O.R.-B. acknowledges support from Fundación Mauricio y Carlota Botton. J.S.P. acknowledges support from the Arnold and Mabel Beckman Foundation via the Arnold O. Beckman Postdoctoral Fellowship in Chemical Sciences.

\bibliography{references}

@article{peter_chirality_2024,
	title = {Chirality dependent photon transport and helical superradiance},
	volume = {6},
	doi = {10.1103/PhysRevResearch.6.023200},
	number = {2},
	journal = {Physical Review Research},
	author = {Peter, Jonah S and Ostermann, Stefan and Yelin, Susanne F},
	year = {2024},
	pages = {023200},
}

@article{peter_chirality-induced_2024,
	title = {Chirality-induced emergent spin-orbit coupling in topological atomic lattices},
	volume = {109},
	issn = {2469-9926, 2469-9934},
	url = {https://link.aps.org/doi/10.1103/PhysRevA.109.043525},
	doi = {10.1103/PhysRevA.109.043525},
	number = {4},
	urldate = {2024-05-20},
	journal = {Physical Review A},
	author = {Peter, Jonah S. and Ostermann, Stefan and Yelin, Susanne F.},
	month = apr,
	year = {2024},
	pages = {043525},
}

@article{dicke_coherence_1954,
	title = {Coherence in {Spontaneous} {Radiation} {Processes}},
	volume = {93},
	issn = {0031-899X},
	url = {https://link.aps.org/doi/10.1103/PhysRev.93.99},
	doi = {10.1103/PhysRev.93.99},
	number = {1},
	urldate = {2024-02-13},
	journal = {Physical Review},
	author = {Dicke, R. H.},
	month = jan,
	year = {1954},
	pages = {99--110},
}

@article{bonifacio_cooperative_1975,
	title = {Cooperative radiation processes in two-level systems: {Superfluorescence}},
	volume = {11},
	copyright = {http://link.aps.org/licenses/aps-default-license},
	issn = {0556-2791},
	shorttitle = {Cooperative radiation processes in two-level systems},
	url = {https://link.aps.org/doi/10.1103/PhysRevA.11.1507},
	doi = {10.1103/PhysRevA.11.1507},
	number = {5},
	urldate = {2025-05-12},
	journal = {Physical Review A},
	author = {Bonifacio, R. and Lugiato, L. A.},
	month = may,
	year = {1975},
	pages = {1507--1521},
}

@article{bonifacio_quantum_1971,
	title = {Quantum {Statistical} {Theory} of {Superradiance}. {I}},
	volume = {4},
	copyright = {http://link.aps.org/licenses/aps-default-license},
	issn = {0556-2791},
	url = {https://link.aps.org/doi/10.1103/PhysRevA.4.302},
	doi = {10.1103/PhysRevA.4.302},
	number = {1},
	urldate = {2025-08-15},
	journal = {Physical Review A},
	author = {Bonifacio, R. and Schwendimann, P. and Haake, Fritz},
	month = jul,
	year = {1971},
	pages = {302--313},
}

@article{St-Jean2017LasingTopologicalEdgeStates,
  author  = {St-Jean, P. and Goblot, V. and Galopin, E. and Lema{\^i}tre, A. and Ozawa, T. and Le Gratiet, L. and Sagnes, I. and Bloch, J. and Amo, A.},
  title   = {Lasing in topological edge states of a one-dimensional lattice},
  journal = {Nature Photonics},
  year    = {2017},
  volume  = {11},
  number  = {10},
  pages   = {651--656},
  doi     = {10.1038/s41566-017-0006-2},
  url     = {https://www.nature.com/articles/s41566-017-0006-2}
}

@article{Baier2016,
  author       = {Baier, Simon and Mark, Manfred J. and Petter, Daniel and Aikawa, Kazuki and Chomaz, Lucas and Cai, Zi and Baranov, Mikhail A. and Zoller, Peter and Ferlaino, Francesca},
  title        = {Extended Bose‑Hubbard Models with Ultracold Magnetic Atoms},
  journal      = {Science},
  volume       = {352},
  number       = {6282},
  pages        = {201--205},
  year         = {2016},
  month        = {apr},
  doi          = {10.1126/science.aac9812},
  url          = {https://doi.org/10.1126/science.aac9812}
}

@article{Norcia2018,
  author       = {Norcia, M. A. and Young, A. W. and Kaufman, A. M.},
  title        = {Microscopic Control and Detection of Ultracold Strontium in Optical‑Tweezer Arrays},
  journal      = {Physical Review X},
  volume       = {8},
  number       = {4},
  eid          = {041054},
  pages        = {041054},
  year         = {2018},
  month        = {dec},
  doi          = {10.1103/PhysRevX.8.041054},
  url          = {https://doi.org/10.1103/PhysRevX.8.041054}
}

@article{Seubert2025,
  author       = {Seubert, Matthias and Hartung, Lukas and Welte, Stephan and Rempe, Gerhard and Distante, Emanuele},
  title        = {Tweezer‑Assisted Subwavelength Positioning of Atomic Arrays in an Optical Cavity},
  journal      = {PRX Quantum},
  volume       = {6},
  number       = {1},
  eid          = {010322},
  pages        = {010322},
  year         = {2025},
  month        = {feb},
  doi          = {10.1103/PRXQuantum.6.010322},
  url          = {https://doi.org/10.1103/PRXQuantum.6.010322}
}

@article{Du2024_dipolar_bilayer,
  author       = {Li Du and Pierre Barral and Michael Cantara and Julius de Hond and Yu‑Kun Lu and Wolfgang Ketterle},
  title        = {Atomic physics on a 50-nm scale: Realization of a bilayer system of dipolar atoms},
  journal      = {Science},
  volume       = {384},
  number       = {6695},
  pages        = {546--551},
  year         = {2024},
  doi          = {10.1126/science.adh3023},
}

@article{glicenstein2020collective,
  title = {Collective Shift in Resonant Light Scattering by a One-Dimensional Atomic Chain},
  author = {Glicenstein, Antoine and Ferioli, Giovanni and \ifmmode \check{S}\else \v{S}\fi{}ibali\ifmmode \acute{c}\else \'{c}\fi{}, Nikola and Brossard, Ludovic and Ferrier-Barbut, Igor and Browaeys, Antoine},
  journal = {Phys. Rev. Lett.},
  volume = {124},
  issue = {25},
  pages = {253602},
  numpages = {6},
  year = {2020},
  month = {Jun},
  publisher = {American Physical Society},
  doi = {10.1103/PhysRevLett.124.253602},
  url = {https://link.aps.org/doi/10.1103/PhysRevLett.124.253602}
}

@article{brow2017optical,
  title = {Optical Control of the Resonant Dipole-Dipole Interaction between Rydberg Atoms},
  author = {de L\'es\'eleuc, Sylvain and Barredo, Daniel and Lienhard, Vincent and Browaeys, Antoine and Lahaye, Thierry},
  journal = {Phys. Rev. Lett.},
  volume = {119},
  issue = {5},
  pages = {053202},
  numpages = {6},
  year = {2017},
  month = {Aug},
  publisher = {American Physical Society},
  doi = {10.1103/PhysRevLett.119.053202},
  url = {https://link.aps.org/doi/10.1103/PhysRevLett.119.053202}
}

@article{Pause2024laser,
author = {Lars Pause and Lukas Sturm and Marcel Mittenb\"{u}hler and Stephan Amann and Tilman Preuschoff and Dominik Sch\"{a}ffner and Malte Schlosser and Gerhard Birkl},
journal = {Optica},
number = {2},
pages = {222--226},
publisher = {Optica Publishing Group},
title = {Supercharged two-dimensional tweezer array with more than 1000 atomic qubits},
volume = {11},
month = {Feb},
year = {2024},
url = {https://opg.optica.org/optica/abstract.cfm?URI=optica-11-2-222},
doi = {10.1364/OPTICA.513551},
}

@misc{holman2024trappingsingleatomsmetasurface,
      title={Trapping of Single Atoms in Metasurface Optical Tweezer Arrays}, 
      author={Aaron Holman and Yuan Xu and Ximo Sun and Jiahao Wu and Mingxuan Wang and Bojeong Seo and Nanfang Yu and Sebastian Will},
      year={2024},
      eprint={2411.05321},
      archivePrefix={arXiv},
      primaryClass={physics.atom-ph},
      url={https://arxiv.org/abs/2411.05321}, 
}

@article{Rui2020,
  author       = {Rui, Jun and Wei, David and Rubio‑Abadal, Antonio and Hollerith, Simon and Zeiher, Johannes and Stamper‑Kurn, Dan M. and Gross, Christian and Bloch, Immanuel},
  title        = {A subradiant optical mirror formed by a single structured atomic layer},
  journal      = {Nature},
  volume       = {583},
  number       = {7816},
  pages        = {369--374},
  year         = {2020},
  month        = {jul},
  doi          = {10.1038/s41586-020-2463-x},
  url          = {https://doi.org/10.1038/s41586-020-2463-x}
}

@phdthesis{Rubies_thesis_mit,
  author       = {Rubies-Bigorda, Oriol},  
  title        = {Light-Induced Collective Interactions in Arrays of Quantum Emitters},  
  school       = {Massachusetts Institute of Technology},
  address      = {Cambridge, MA},
  year         = {2025},  
  type         = {PhD thesis},
  url          = {https://dspace.mit.edu/handle/1721.1/164157}
}

@misc{Greiner_superradiance,
      title={Many-Body Super- and Subradiance in Ordered Atomic Arrays}, 
      author={Alec Douglas and Lin Su and Michal Szurek and Robin Groth and Sandra Brandstetter and Ognjen Markovic and Oriol Rubies-Bigorda and Stefan Ostermann and Susanne F. Yelin and Markus Greiner},
      year={2026},
      eprint={2604.11795},
      archivePrefix={arXiv},
      primaryClass={quant-ph},
      url={https://arxiv.org/abs/2604.11795}, 
}

@article{chang2013selforganization,
  title = {Self-Organization of Atoms along a Nanophotonic Waveguide},
  author = {Chang, D. E. and Cirac, J. I. and Kimble, H. J.},
  journal = {Phys. Rev. Lett.},
  volume = {110},
  issue = {11},
  pages = {113606},
  numpages = {6},
  year = {2013},
  month = {Mar},
  publisher = {American Physical Society},
  doi = {10.1103/PhysRevLett.110.113606},
  url = {https://link.aps.org/doi/10.1103/PhysRevLett.110.113606}
}

@article{Dalibard_1985,
doi = {10.1088/0022-3700/18/8/019},
url = {https://dx.doi.org/10.1088/0022-3700/18/8/019},
year = {1985},
month = {apr},
publisher = {},
volume = {18},
number = {8},
pages = {1661},
author = {J Dalibard and C Cohen-Tannoudji},
title = {Atomic motion in laser light: connection between semiclassical and quantum descriptions},
journal = {Journal of Physics B: Atomic and Molecular Physics}
}

@article{WALLIS1995203,
title = {Quantum theory of atomic motion in laser light},
journal = {Physics Reports},
volume = {255},
number = {4},
pages = {203-287},
year = {1995},
issn = {0370-1573},
doi = {https://doi.org/10.1016/0370-1573(94)00090-P},
url = {https://www.sciencedirect.com/science/article/pii/037015739400090P},
author = {H. Wallis}
}

@article{Chu1985,
  title = {Three-dimensional viscous confinement and cooling of atoms by resonance radiation pressure},
  author = {Chu, Steven and Hollberg, L. and Bjorkholm, J. E. and Cable, Alex and Ashkin, A.},
  journal = {Phys. Rev. Lett.},
  volume = {55},
  issue = {1},
  pages = {48--51},
  numpages = {0},
  year = {1985},
  month = {Jul},
  publisher = {American Physical Society},
  doi = {10.1103/PhysRevLett.55.48},
  url = {https://link.aps.org/doi/10.1103/PhysRevLett.55.48}
}

@article{Dalibard1989,
author = {J. Dalibard and C. Cohen-Tannoudji},
journal = {J. Opt. Soc. Am. B},
number = {11},
pages = {2023--2045},
publisher = {Optica Publishing Group},
title = {Laser cooling below the Doppler limit by polarization gradients: simple theoretical models},
volume = {6},
month = {Nov},
year = {1989},
url = {https://opg.optica.org/josab/abstract.cfm?URI=josab-6-11-2023},
doi = {10.1364/JOSAB.6.002023},
}

@article{Domokos2002,
  title={Collective cooling and self-organization of atoms in a cavity},
  author={Domokos, Peter and Ritsch, Helmut},
  journal={Physical Review Letters},
  volume={89},
  number={25},
  pages={253003},
  year={2002}
}

@article{Black2003,
  title = {Observation of Collective Friction Forces due to Spatial Self-Organization of Atoms: From Rayleigh to Bragg Scattering},
  author = {Black, Adam T. and Chan, Hilton W. and Vuleti\ifmmode \acute{c}\else \'{c}\fi{}, Vladan},
  journal = {Phys. Rev. Lett.},
  volume = {91},
  issue = {20},
  pages = {203001},
  numpages = {4},
  year = {2003},
  month = {Nov},
  publisher = {American Physical Society},
  doi = {10.1103/PhysRevLett.91.203001},
  url = {https://link.aps.org/doi/10.1103/PhysRevLett.91.203001}
}

@article{Baumann2010,
  title={Dicke quantum phase transition with a superfluid gas in an optical cavity},
  author={Baumann, K and Guerlin, C and Brennecke, F and Esslinger, T},
  journal={Nature},
  volume={464},
  pages={1301--1306},
  year={2010},
  doi= {https://doi.org/10.1038/nature09009}
}

@misc{SteckRb87,
  author       = {Daniel A. Steck},
  title        = {{Rubidium 87 D Line Data}},
  year         = {2019},
  howpublished = {\url{http://steck.us/alkalidata/rubidium87numbers.pdf}},
  note         = {Revision 2.1.6},
}

@article{BlushSrDipole,
  author       = {K. Blush and M. Auzinsh},
  title        = {Calculation of level-crossing signals for spin-1 and spin-2 systems using irreducible tensor operators},
  journal      = {Journal of the Optical Society of America B},
  year         = {2003},
  volume       = {20},
  number       = {5},
  pages        = {103--111},
  doi          = {10.1364/JOSAB.20.000103}
}

@article{SantraSrNarrow,
  author       = {R. Santra and E. Arimondo and T. Ido and C. H. Greene and J. Ye},
  title        = {High-accuracy optical clock via three-level coherence in neutral bosonic $^{88}$Sr},
  journal      = {Physical Review A},
  year         = {2004},
  volume       = {69},
  pages        = {042510},
  doi          = {10.1103/PhysRevA.69.042510}
}

@article{TakasuYbNarrow,
  author       = {Y. Takasu and K. Maki and K. Komori and T. Takano and K. Honda and M. Kumakura and T. Yabuzaki and Y. Takahashi},
  title        = {Spin-Singlet Bose-Einstein Condensate of Two-Electron Atoms},
  journal      = {Physical Review Letters},
  year         = {2003},
  volume       = {91},
  pages        = {040404},
  doi          = {10.1103/PhysRevLett.91.040404}
}

@article{PorsevYbData,
  author       = {S. G. Porsev and A. Derevianko},
  title        = {Multipolar theory of blackbody radiation shift of atomic energy levels and its implications for optical lattice clocks},
  journal      = {Physical Review A},
  year         = {2004},
  volume       = {69},
  pages        = {021403},
  doi          = {10.1103/PhysRevA.69.021403}
}

@article{shahmoon2020quantum,
  title = {Quantum optomechanics of a two-dimensional atomic array},
  author = {Shahmoon, Ephraim and Lukin, Mikhail D. and Yelin, Susanne F.},
  journal = {Phys. Rev. A},
  volume = {101},
  issue = {6},
  pages = {063833},
  numpages = {13},
  year = {2020},
  month = {Jun},
  publisher = {American Physical Society},
  doi = {10.1103/PhysRevA.101.063833},
  url = {https://link.aps.org/doi/10.1103/PhysRevA.101.063833}
}

@incollection{SHAHMOON20191,
title = {Chapter One - Collective motion of an atom array under laser illumination},
editor = {Louis F. Dimauro and Hélène Perrin and Susanne F. Yelin},
series = {Advances In Atomic, Molecular, and Optical Physics},
publisher = {Academic Press},
volume = {68},
pages = {1-38},
year = {2019},
issn = {1049-250X},
doi = {https://doi.org/10.1016/bs.aamop.2019.03.001},
url = {https://www.sciencedirect.com/science/article/pii/S1049250X19300011},
author = {Ephraim Shahmoon and Mikhail D. Lukin and Susanne F. Yelin}
}

@article{maximo2018optical,
  title = {Optical binding with cold atoms},
  author = {M\'aximo, C. E. and Bachelard, R. and Kaiser, R.},
  journal = {Phys. Rev. A},
  volume = {97},
  issue = {4},
  pages = {043845},
  numpages = {7},
  year = {2018},
  month = {Apr},
  publisher = {American Physical Society},
  doi = {10.1103/PhysRevA.97.043845},
  url = {https://link.aps.org/doi/10.1103/PhysRevA.97.043845}
}

@article{BuckleyBonanno2024,
  author       = {Samuel Buckley-Bonanno and Stefan Ostermann and Yidan Wang and Susanne F. Yelin},
  title        = {Two-dimensional motion of an impurity under dynamic light-induced dipole forces in an atomic subwavelength array},
  journal      = {arXiv preprint arXiv:2407.11113},
  year         = {2024},
  url          = {https://arxiv.org/abs/2407.11113},
  note         = {11 pages, 7 figures},
  eprint       = {2407.11113},
  archivePrefix= {arXiv},
  primaryClass = {quant-ph}
}

@article{HUANG2023100470,
title = {Metasurface holographic optical traps for ultracold atoms},
journal = {Progress in Quantum Electronics},
volume = {89},
pages = {100470},
year = {2023},
issn = {0079-6727},
doi = {https://doi.org/10.1016/j.pquantelec.2023.100470},
url = {https://www.sciencedirect.com/science/article/pii/S0079672723000198},
author = {Xiaoyan Huang and Weijun Yuan and Aaron Holman and Minho Kwon and Stuart J. Masson and Ricardo Gutierrez-Jauregui and Ana Asenjo-Garcia and Sebastian Will and Nanfang Yu}
}

@article{asboth2005self,
  title = {Self-organization of atoms in a cavity field: Threshold, bistability, and scaling laws},
  author = {Asb\'oth, J. K. and Domokos, P. and Ritsch, H. and Vukics, A.},
  journal = {Phys. Rev. A},
  volume = {72},
  issue = {5},
  pages = {053417},
  numpages = {12},
  year = {2005},
  month = {Nov},
  publisher = {American Physical Society},
  doi = {10.1103/PhysRevA.72.053417},
  url = {https://link.aps.org/doi/10.1103/PhysRevA.72.053417}
}

@article{Barredo2016,
  title = {An atom-by-atom assembler of defect-free arbitrary 2D atomic arrays},
  author = {Barredo, Daniel and de Léséleuc, Sylvain and Lienhard, Vincent and Lahaye, Thierry and Browaeys, Antoine},
  journal = {Science},
  volume = {354},
  number = {6315},
  pages = {1021--1023},
  year = {2016},
  doi = {10.1126/science.aah3778},
  url = {https://www.science.org/doi/10.1126/science.aah3778}
}

@article{Endres2016,
  title = {Atom-by-atom assembly of defect-free one-dimensional cold atom arrays},
  author = {Endres, Manuel and Bernien, Hannes and Keesling, Alexander and Levine, Harry and Anschuetz, Eric R. and Krajenbrink, Alexandre and Senko, Crystal and Vuletić, Vladan and Greiner, Markus and Lukin, Mikhail D.},
  journal = {Science},
  volume = {354},
  number = {6315},
  pages = {1024--1027},
  year = {2016},
  doi = {10.1126/science.aah3752},
  url = {https://www.science.org/doi/10.1126/science.aah3752}
}

@article{Koepke1996,
  title = {The crystal structure of the light-harvesting complex II (B800–850) from Rhodospirillum molischianum},
  author = {Koepke, J. and Hu, X. and Muenke, C. and Schulten, K. and Michel, H.},
  journal = {Structure},
  volume = {4},
  number = {5},
  pages = {581--597},
  year = {1996},
  doi = {10.1016/S0969-2126(96)00063-9},
  url = {https://www.sciencedirect.com/science/article/pii/S0969212696000639}
}

@article{schutz2015thermodynamics,
  title = {Thermodynamics and dynamics of atomic self-organization in an optical cavity},
  author = {Sch\"utz, Stefan and J\"ager, Simon B. and Morigi, Giovanna},
  journal = {Phys. Rev. A},
  volume = {92},
  issue = {6},
  pages = {063808},
  numpages = {18},
  year = {2015},
  month = {Dec},
  publisher = {American Physical Society},
  doi = {10.1103/PhysRevA.92.063808},
  url = {https://link.aps.org/doi/10.1103/PhysRevA.92.063808}
}

@article{black2003observation,
  title = {Observation of Collective Friction Forces due to Spatial Self-Organization of Atoms: From Rayleigh to Bragg Scattering},
  author = {Black, Adam T. and Chan, Hilton W. and Vuleti\ifmmode \acute{c}\else \'{c}\fi{}, Vladan},
  journal = {Phys. Rev. Lett.},
  volume = {91},
  issue = {20},
  pages = {203001},
  numpages = {4},
  year = {2003},
  month = {Nov},
  publisher = {American Physical Society},
  doi = {10.1103/PhysRevLett.91.203001},
  url = {https://link.aps.org/doi/10.1103/PhysRevLett.91.203001}
}

@article{su1979solitons,
  title = {Solitons in Polyacetylene},
  author = {Su, W. P. and Schrieffer, J. R. and Heeger, A. J.},
  journal = {Phys. Rev. Lett.},
  volume = {42},
  issue = {25},
  pages = {1698--1701},
  numpages = {0},
  year = {1979},
  month = {Jun},
  publisher = {American Physical Society},
  doi = {10.1103/PhysRevLett.42.1698},
  url = {https://link.aps.org/doi/10.1103/PhysRevLett.42.1698}
}

@article{Ritsch_subradiance_9fold,
author = {Maria Moreno-Cardoner and Raphael Holzinger and Helmut Ritsch},
journal = {Opt. Express},
number = {7},
pages = {10779--10791},
publisher = {Optica Publishing Group},
title = {Efficient nano-photonic antennas based on dark states in quantum emitter rings},
volume = {30},
month = {Mar},
year = {2022},
url = {https://opg.optica.org/oe/abstract.cfm?URI=oe-30-7-10779},
doi = {10.1364/OE.437396},
}

@article{Domokos2003,
  author    = {Peter Domokos and Helmut Ritsch},
  title     = {Mechanical effects of light in optical resonators},
  journal   = {Journal of the Optical Society of America B},
  volume    = {20},
  number    = {5},
  pages     = {1098--1130},
  year      = {2003},
  doi       = {10.1364/JOSAB.20.001098},
  url       = {https://opg.optica.org/josab/abstract.cfm?uri=josab-20-5-1098}
}

@article{eldredge2016self,
  author       = {Zachary Eldredge and Pablo Solano and Darrick Chang and Alexey V. Gorshkov},
  title        = {Self-organization of atoms coupled to a chiral reservoir},
  journal      = {Physical Review A},
  volume       = {94},
  number       = {5},
  pages        = {053855},
  year         = {2016},
  doi          = {10.1103/PhysRevA.94.053855},
  url          = {https://link.aps.org/doi/10.1103/PhysRevA.94.053855}
}

@article{ho2025optomechanical,
  author       = {Jacquelyn Ho and Yue-Hui Lu and Tai Xiang and Cosimo C. Rusconi and Stuart J. Masson and Ana Asenjo-Garcia and Zhenjie Yan and Dan M. Stamper-Kurn},
  title        = {Optomechanical self-organization in a mesoscopic atom array},
  journal      = {Nature Physics},
  year         = {2025},
  doi          = {10.1038/s41567-025-02916-7},
  url          = {https://www.nature.com/articles/s41567-025-02916-7}
}

@article{AsenjoGarcia2017,
  title = {Exponential Improvement in Photon Storage Fidelities Using Subradiance and ``Selective Radiance'' in Atomic Arrays},
  author = {Asenjo-Garcia, A. and Moreno-Cardoner, M. and Albrecht, A. and Kimble, H. J. and Chang, D. E.},
  journal = {Phys. Rev. X},
  volume = {7},
  issue = {3},
  pages = {031024},
  numpages = {36},
  year = {2017},
  month = {Aug},
  publisher = {American Physical Society},
  doi = {10.1103/PhysRevX.7.031024},
  url = {https://link.aps.org/doi/10.1103/PhysRevX.7.031024}
}

@article{harochesuperradiance,
  title = {Superradiance: An essay on the theory of collective
spontaneous emission},
  author = { M. Gross and S. Haroche},
  journal = {Physics Reports},
  volume = {93},
  issue = {5},
  pages = {301 – 396},
  numpages = {6},
  year = {1982},
  month = {Sep},
  publisher = {American Physical Society},
  doi = {10.1103/PhysRevLett.110.113606},
  url = {https://link.aps.org/doi/10.1103/PhysRevLett.110.113606}
}

@article{ipr,
  title = {Non-Hermitian Euclidean random matrix theory},
  author = {Goetschy, A. and Skipetrov, S. E.},
  journal = {Phys. Rev. E},
  volume = {84},
  issue = {1},
  pages = {011150},
  numpages = {10},
  year = {2011},
  month = {Jul},
  publisher = {American Physical Society},
  doi = {10.1103/PhysRevE.84.011150},
  url = {https://link.aps.org/doi/10.1103/PhysRevE.84.011150}
}

@article{nonhermtop,
  title = {Topological photonic states in one-dimensional dimerized ultracold atomic chains},
  author = {Wang, B. X. and Zhao, C. Y.},
  journal = {Phys. Rev. A},
  volume = {98},
  issue = {2},
  pages = {023808},
  numpages = {12},
  year = {2018},
  month = {Aug},
  publisher = {American Physical Society},
  doi = {10.1103/PhysRevA.98.023808},
  url = {https://link.aps.org/doi/10.1103/PhysRevA.98.023808}
}

@article{ring_raphael,
author = {Raphael Holzinger and Jonah S. Peter and Stefan Ostermann and Helmut Ritsch and Susanne Yelin},
journal = {Optica Quantum},
number = {2},
pages = {57--63},
publisher = {Optica Publishing Group},
title = {Harnessing quantum emitter rings for efficient energy transport and trapping},
volume = {2},
month = {Apr},
year = {2024},
url = {https://opg.optica.org/opticaq/abstract.cfm?URI=opticaq-2-2-57},
doi = {10.1364/OPTICAQ.510021}
}

@article{oriolraphaelcooling,
  title = {Collectively enhanced ground-state cooling in subwavelength atomic arrays},
  author = {Rubies-Bigorda, Oriol and Holzinger, Raphael and Asenjo-Garcia, Ana and Romero-Isart, Oriol and Ritsch, Helmut and Ostermann, Stefan and Gonzalez-Ballestero, Carlos and Yelin, Susanne F. and Rusconi, Cosimo C.},
  journal = {Phys. Rev. A},
  volume = {112},
  issue = {2},
  pages = {023714},
  numpages = {24},
  year = {2025},
  month = {Aug},
  publisher = {American Physical Society},
  doi = {10.1103/bhwv-ndtj},
  url = {https://link.aps.org/doi/10.1103/bhwv-ndtj}
}

@article{Perczel_topology,
  title = {Topological Quantum Optics in Two-Dimensional Atomic Arrays},
  author = {Perczel, J. and Borregaard, J. and Chang, D. E. and Pichler, H. and Yelin, S. F. and Zoller, P. and Lukin, M. D.},
  journal = {Phys. Rev. Lett.},
  volume = {119},
  issue = {2},
  pages = {023603},
  numpages = {6},
  year = {2017},
  month = {Jul},
  publisher = {American Physical Society},
  doi = {10.1103/PhysRevLett.119.023603},
  url = {https://link.aps.org/doi/10.1103/PhysRevLett.119.023603}
}

@article{lehmberg,
  title = {Radiation from an $N$-Atom System. I. General Formalism},
  author = {Lehmberg, R. H.},
  journal = {Phys. Rev. A},
  volume = {2},
  issue = {3},
  pages = {883--888},
  numpages = {0},
  year = {1970},
  month = {Sep},
  publisher = {American Physical Society},
  doi = {10.1103/PhysRevA.2.883},
  url = {https://link.aps.org/doi/10.1103/PhysRevA.2.883}
}

\newpage

 \onecolumngrid
 \appendix

\section{Electromagnetic Green's Tensor in Free Space}
\label{app:greentensor}
The coherent and dissipative dipole-dipole couplings between emitters $n$ and $m$ read
\begin{align} \label{3d-dipole}
   J_{nm} - \frac{i\Gamma_{nm}}{2} = -\frac{3 \pi \Gamma_0}{\omega_0} \mathbf{d}^\dagger \cdot \mathbf{G}(\mathbf{r}_{nm},\omega_0) \cdot \mathbf{d},
\end{align}
where $\mathbf{d}$ is the transition dipole moment matrix element, $\mathbf{r}_{nm} = \mathbf{r}_n-\mathbf{r}_m$ is the connecting vector between emitters $n$ and $m$. The Green's tensor $\mathbf{G}(\mathbf{r}_{nm},\omega_0)$ is the propagator of the electromagnetic
field between emitter positions $\boldsymbol{r}_n$ and $\boldsymbol{r}_m$, and reads for 3D dipole-dipole interactions~\cite{lehmberg}
\begin{align}
\mathbf{G}(\mathbf{r}_{nm},\omega_0) = \frac{e^{i k_0 r_{nm}}}{4\pi k_0^2 r_{nm}^3} \bigg[ \left( k_0^2 r_{nm}^2 + ik_0 r_{nm} -1 \right) \mathbbm{1}  +  \left(-k_0^2 r_{nm}^2 - 3i k_0 r_{nm} + 3 \right) \frac{\mathbf{r}_{nm} \otimes \mathbf{r}_{nm}}{r_{nm}^2} \bigg],
\end{align}
with $r_{nm} = |\mathbf{r}_{nm}|$ and $k_0 = 2\pi/\lambda_0$, where $\lambda_0$ is the wavelength of light emitted by the emitters. The spontaneous decay rate is equal for all atoms, $\Gamma_{nn} = \Gamma_0$, while the self-energy term $J_{nn}$ is set to zero, as it simply leads to renormalization of the transition energies $\omega_0$.

\section{Validation of the Weak Driving Limit and Adiabatic Approximation}
\label{app:refdim}
\begin{figure}[hbt]
\includegraphics[width=\linewidth]{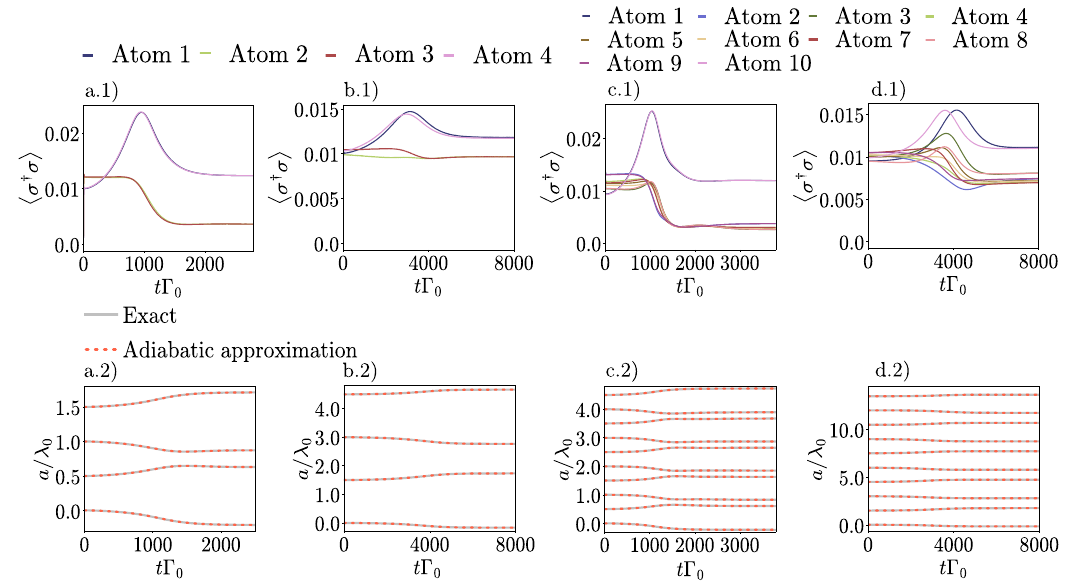}
\caption{\label{fig:apendixb} Self-organization in an atomic chain showing the excited state populations (top row) and atomic positions (bottom row). Dynamics are shown with initial position disorder within the range $[-0.01\,a_0,\; 0.01\,a_0]$. Parameters for all panels: $\omega=1.0\omega_r$, $\Omega=0.05\Gamma_0$, $\delta=0$, and same laser orientation as in the main text. Additional parameters for panels a) $N=4$, $a_0=0.5$ b) $N=4$, $a_0=1.5$ c) $N=10$, $a_0=0.5$ d) $N=10$, $a_0=1.5$.}
\end{figure}

In the main text, we assumed the weak driving limit so that the atomic coherences can be evaluated in terms of their mean values. This approximation is warranted provided the excited state populations remain small throughout the time evolution. The top row of Fig.~\ref{fig:apendixb} shows the excited state populations as the atomic positions evolve towards a self-organized structure. The populations reach a maximum of about $ \langle\tilde{\sigma}^{\dagger}\tilde{\sigma}\rangle\approx0.025$, thus confirming the validity of the mean field approximation. 

In deriving the effective potential for $N=2$, we performed an adiabatic approximation by neglecting the time derivative of the atomic coherences $\langle\dot{\tilde{\sigma}}\rangle$. This simplifies the equations of motion by reducing the problem to solving a set of $2N$ instead of $3N$ differential equations, improving computational efficiency. The bottom row of Fig.~\ref{fig:apendixb} shows the time evolution of the atomic positions in the linear chain for $N=4$ (panels a.2,b.2) and $N=10$ (panels c.2,d.2). As shown by the overlapping curves, the adiabatic and exact solutions are in good agreement, even for larger system sizes. 

Regarding the time required to reach the self-organized structure: in our approach the transient evolution primarily serves as a numerical relaxation toward a stationary configuration, and the convergence time depends on the chosen auxiliary damping parameter $\gamma$ as well as on the local curvature of the effective potential. For this reason, the integration time in units of $\Gamma_0^{-1}$ should be interpreted as an algorithmic convergence time rather than as a quantitative prediction of experimental self-organization dynamics. However, it is clear that the steady-state will be reached faster with higher drives, but we are limited on the value of $\Omega$ by the single-excitation regime.

\section{Disordered Configurations and Non-zero Detunings}
\label{app:others}
\begin{figure}[htbp]
\centering
\includegraphics[width=0.7\textwidth]{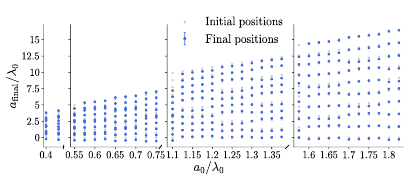}
\caption{\label{fig:x_others} Initial (grey dots) and final positions (blue dots) for the self-organized configurations corresponding to the grey regions in Fig.~\ref{fig:dim}a). The final positions are an average of 20 runs for initial uniform positional disorder within the range $[-0.01\,a_0,\; 0.01\,a_0]$.}
\end{figure}

In the main text, we presented the dimer strength for chains comprised of two dominant, consecutive alternating distances. However, as seen indicated by the shaded regions in Fig.~\ref{fig:dim}b), larger system sizes admit configurations where no such pattern is seen. These configurations are shown in Fig.~\ref{fig:x_others} for a variety of disordered initial positions. At small initial distances, dimers still form, though less frequently as $a_0$ increases. Starting from $a_0=0.775\lambda_0$, the system transitions to the ordered chain configurations shown in the main text. As $a_0$ increases further, the periodic order is lost and the chains instead form aggregations of atoms that emerge even in the absence of positional disorder.

\begin{figure}[hbt]
\includegraphics[width=\linewidth]{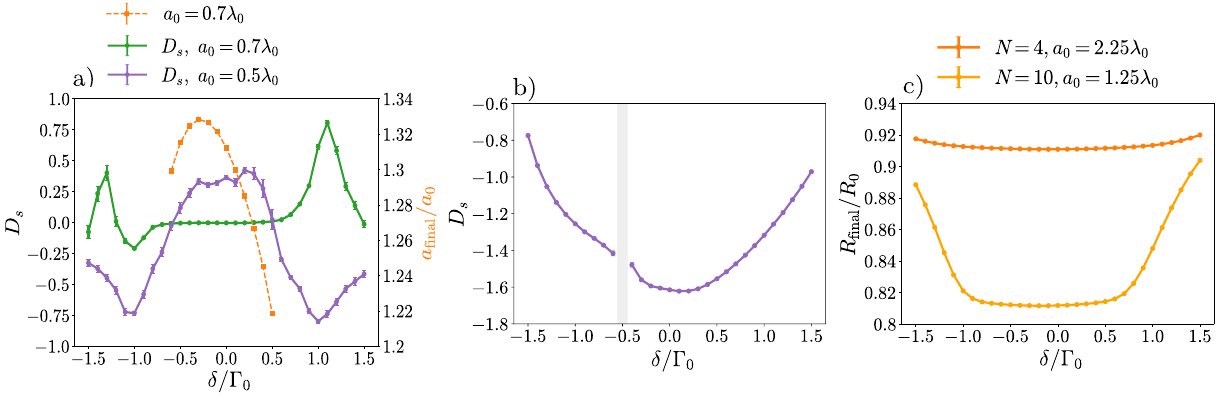}
\caption{\label{fig:scandetchain} Dimer strengths and final lattice constants as a function of laser detuning. Panels a) ($N=4$) and b) ($N=10$) correspond to chains with $\omega=\omega_r$. The shaded region indicates no steady-state solution was found. c) Final radius of ring geometries with $N=4$ (red) and $N=10$ (orange). In both cases, $\omega=0.1\omega_r$. All results are averaged over 20 runs for initial uniform positional disorder within the range $[-0.01\,a_0,\; 0.01\,a_0]$.. Error bars denote standard deviation and are in general smaller than the symbol size.}
\end{figure}

Figs.~\ref{fig:scandetchain}a) ($N=4$) and b) ($N=10$) show the dimer strength and lattice constant ratio for chains driven by lasers with non-zero detunings. For $N=10$ and $\delta=-0.5\Gamma_0$, the shaded region indicates that no steady-state is reached for the parameters considered. The scan for the ring geometry is shown in Fig.~\ref{fig:scandetchain}c). By contrast, the simulation time increases away from resonance, as the atoms take longer to reach a steady-state.

\section{Lower Limit of Trap Frequency due to Zero-Point Motion}
\label{app:zpm}
In all our simulations, the capacity for the initial system to self-organize depends strongly on the depths of the harmonic traps. At the same time, we also require that the lattice spacings are similar to or smaller than the wavelength of resonant atomic transition $\lambda_0$ in order to observe significant dipole-dipole interactions. To ensure that atomic zero-point motion does not lead to significant overlap between neighboring sites in an optical lattice, it is required that the spatial extent of the ground-state wavefunction be much smaller than the lattice spacing. Approximating each site as a (two-dimensional) harmonic oscillator, the zero-point motion is given by \( x_\mathrm{zpm} = \sqrt{\hbar / (2 m \omega)} \), where \( \omega \) is the trap frequency and \( m \) is the atomic mass. For a lattice spacing \( a \), we require \( x_\mathrm{zpm}  \ll a \), leading to the condition
\[
\frac{\omega}{\Gamma_0} \gg \frac{\hbar}{2 m a^2 \Gamma_0},
\]
where \( \Gamma_0 \) is the natural atomic linewidth. In the tables below, we list the relevant parameters for the atomic species Rubidium~\cite{SteckRb87}, Strontium~\cite{BlushSrDipole,SantraSrNarrow} and Ytterbium~\cite{TakasuYbNarrow,PorsevYbData}, commonly employed in quantum optical platforms. These considerations demonstrate that for broad optical transitions (e.g., Rb D2, Sr and Yb dipole lines), even shallow traps would easily satisfy the localization condition. However, narrow line transitions such as the Sr \(689\,\mathrm{nm}\) and Yb \(556\,\mathrm{nm}\) lines require significantly tighter confinement, especially for sub-wavelength lattices (\( a \ll \lambda_0 \)), in order to suppress wavefunction overlap and ensure validity of semiclassical treatment of motion. 

\begin{table}[ht!]
\setlength{\tabcolsep}{14pt}  
\centering
\caption{\textbf{Atomic parameters for relevant transitions}}
\renewcommand{\arraystretch}{1.3}
\begin{tabular}{lccc}
\toprule
\textbf{Atom / Transition} & \boldmath$\lambda_0$ [nm] & \boldmath$\Gamma_0 / 2\pi$ [MHz] & \boldmath$m$ [kg] \\
\midrule
$^{87}$Rb D2 ($5^2S_{1/2} \rightarrow 5^2P_{3/2}$) & 780 & 6.07 & $1.44 \times 10^{-25}$ \\
$^{88}$Sr Dipole ($^1S_0 \rightarrow ^1P_1$) & 461 & 32 & $1.46 \times 10^{-25}$ \\
$^{88}$Sr Narrow ($^1S_0 \rightarrow ^3P_1$) & 689 & 0.0075 & $1.46 \times 10^{-25}$ \\
$^{174}$Yb Dipole ($^1S_0 \rightarrow ^1P_1$) & 399 & 29 & $2.89 \times 10^{-25}$ \\
$^{174}$Yb Narrow ($^1S_0 \rightarrow ^3P_1$) & 556 & 0.182 & $2.89 \times 10^{-25}$ \\
\bottomrule
\end{tabular}
\end{table}

\begin{table}[h]
\setlength{\tabcolsep}{14pt}  
\centering
\caption{\textbf{Minimum required trap frequency ratio \boldmath$\omega/\Gamma_0$ for various spacings}}
\renewcommand{\arraystretch}{1.3}
\begin{tabular}{lccc}
\toprule
\textbf{Atom / Transition} & $a=1.5\lambda_0$ & $a=1.0\lambda_0$ & $a=0.5\lambda_0$ \\
\midrule
$^{87}$Rb D2 &  $7.6 \times 10^{-6}$ & $1.7 \times 10^{-5}$ & $6.8 \times 10^{-5}$ \\
$^{88}$Sr Dipole &  $8.9 \times 10^{-6}$ & $2.0 \times 10^{-5}$ & $8.0 \times 10^{-5}$ \\
$^{88}$Sr Narrow  & $1.8 \times 10^{-2}$ & $4.0 \times 10^{-2}$ & $1.6 \times 10^{-1}$ \\
$^{174}$Yb Dipole  & $2.9 \times 10^{-6}$ & $6.5 \times 10^{-6}$ & $2.6 \times 10^{-5}$ \\
$^{174}$Yb Narrow  & $5.6 \times 10^{-4}$ & $1.3 \times 10^{-3}$ & $5.2 \times 10^{-3}$ \\
\bottomrule
\end{tabular}
\end{table}

\end{document}